\documentclass[conference, 10pt]{IEEEtran}

\usepackage[utf8]{inputenc}
\usepackage{graphicx}
\usepackage{subfig}
\usepackage{xcolor}
\usepackage{amsmath,amssymb}
\usepackage{ifthen}
\usepackage{xspace}

\usepackage{algorithm}
\usepackage{algpseudocode}
\usepackage{color}
\usepackage{listings}
\usepackage{pifont}
\usepackage{multirow}
\usepackage{url}
\usepackage[T1]{fontenc}
\usepackage{textcomp}
\usepackage{balance}
\usepackage{tikz}

\usepackage{listings}

\colorlet{punct}{red!60!black}
\definecolor{background}{HTML}{EEEEEE}
\definecolor{delim}{RGB}{20,105,176}
\colorlet{numb}{magenta!60!black}

\lstdefinelanguage{json}{
    basicstyle=\normalfont\ttfamily,
    numbers=left,
    numberstyle=\scriptsize,
    stepnumber=1,
    numbersep=8pt,
    showstringspaces=false,
    breaklines=true,
    frame=lines,
    backgroundcolor=\color{background},
    literate=
     *{0}{{{\color{numb}0}}}{1}
      {1}{{{\color{numb}1}}}{1}
      {2}{{{\color{numb}2}}}{1}
      {3}{{{\color{numb}3}}}{1}
      {4}{{{\color{numb}4}}}{1}
      {5}{{{\color{numb}5}}}{1}
      {6}{{{\color{numb}6}}}{1}
      {7}{{{\color{numb}7}}}{1}
      {8}{{{\color{numb}8}}}{1}
      {9}{{{\color{numb}9}}}{1}
      {:}{{{\color{punct}{:}}}}{1}
      {,}{{{\color{punct}{,}}}}{1}
      {\{}{{{\color{delim}{\{}}}}{1}
      {\}}{{{\color{delim}{\}}}}}{1}
      {[}{{{\color{delim}{[}}}}{1}
      {]}{{{\color{delim}{]}}}}{1},
}

\newboolean{showcomments}
\setboolean{showcomments}{false}
\ifthenelse{\boolean{showcomments}}
{ \newcommand{\mynote}[3]{
   \fbox{\bfseries\sffamily\scriptsize#1}
   {\small$\blacktriangleright$\textsf{\emph{\color{#3}{#2}}}$\blacktriangleleft$}}}
{ \newcommand{\mynote}[3]{}}

\newcommand{\projecttitle}{\textsc{PubSub-SGX}\xspace}
\newcommand{\sys}{\textsc{PubSub-SGX}\xspace}

\newcommand{\eupro}{\textsc{SELIS}\xspace}

\definecolor{darkgreen}{rgb}{0.3,0.5,0.3}
\definecolor{darkblue}{rgb}{0.3,0.3,0.5}
\definecolor{darkred}{rgb}{0.5,0.3,0.3}

\begin{document}

\title{\Large PubSub-SGX: Exploiting Trusted Execution Environments\\for Privacy-Preserving Publish/Subscribe Systems}

\author{\IEEEauthorblockN{Sergei Arnautov\IEEEauthorrefmark{1}, Andrey Brito\IEEEauthorrefmark{3}, Pascal Felber\IEEEauthorrefmark{2}, Christof Fetzer\IEEEauthorrefmark{1}, Franz Gregor\IEEEauthorrefmark{1}, Robert Krahn\IEEEauthorrefmark{1}, \\Wojciech Ozga\IEEEauthorrefmark{1}, André Martin\IEEEauthorrefmark{1}, Valerio Schiavoni\IEEEauthorrefmark{2}, Fábio Silva\IEEEauthorrefmark{3}, Marcus Tenorio\IEEEauthorrefmark{3} and Nikolaus Thümmel\IEEEauthorrefmark{1}}
\\
\IEEEauthorblockA{\textit{TU Dresden\IEEEauthorrefmark{1}} - \textit{Université de Neuchâtel}\IEEEauthorrefmark{2} - \textit{Universidade Federal de Campina Grande}\IEEEauthorrefmark{3}\\
Dresden, Germany - Neuchâtel, Switzerland - Campina Grande, PB, Brazil\vspace{-2.5ex}}}

\maketitle

\begin{abstract}
This paper presents \projecttitle, a content-based publish-subscribe system that exploits trusted execution environments (TEEs), such as Intel SGX, to guarantee confidentiality and integrity of data as well as anonymity and privacy of publishers and subscribers.
We describe the technical details of our Python implementation, as well as the
 required system support introduced to deploy our system in a container-based runtime. 
Our evaluation results show that our approach is sound, while at the same time highlighting the performance and scalability trade-offs. 
In particular, by supporting \emph{just-in-time compilation} inside of TEEs, Python programs inside of TEEs are in general faster than when executed natively using  standard CPython.
\end{abstract}

\section{Introduction}\label{sec:introduction}

During the past decade, we have been witnessing a continuous growth and adoption of new information technology in various business areas.
Along with this adoption process, old-fashioned paper workflows as well as
 obsolete communication channels are continuously replaced by modern information systems for data and information exchange.

The publish/subscribe paradigm~\cite{eugster2003many} is a key technology, as it enables large groups of participants and systems to exchange information in an efficient manner.
In contrast to the traditional point-to-point communication, the paradigm comes with the benefit of decoupling data producers (publishers) from the consumers (subscribers).
The decoupling occurs in two dimensions: time and space.
Neither publishers nor subscribers are aware of each other.
Instead, they communicate only indirectly through the publish/subscribe system.
It is the system's responsibility to route messages issued by a publisher to subscribers that previously expressed their interest.

The interest for certain types of messages is typically expressed using subscriptions.
Subscriptions can be arbitrarily formed based on the filtering and routing capabilities of the publish/subscribe system.
The simplest form of publish/subscribe systems are topic-based~\cite{castro2002scribe,zhuang2001bayeux,ramasubramanian2006corona}, where publishers issue messages on a specific topic, while subscribers register to the topics they are interesting in.

While the topic-based publish/subscribe model suffices for most application scenarios, some still require more sophisticated filtering techniques, e.g., taking into account the message content.
Consider for example a truck fleet of a freight forwarder that continuously publishes its geo-location information using the publish/subscribe system~\cite{trossen2010not}.
A topic-based publish/subscribe system would be sufficient if subscribers are only interested in receiving updates from specific trucks identified through some unique identifier.
However, applications originating from the logistics domain have often more complex requirements~\cite{liang2010real} such as receiving GPS updates only if trucks are approaching the destination,
i.e., entering a bounding box where the longitude and latitude information stored in the message body needs to be taken into account for routing.

In its original design, the publish/subscribe paradigm provides complete time and space decoupling, i.e., the anonymity of publishers as well as subscribers.
While this works well for many classical applications scenarios, such as a stock exchange~\cite{Barazzutti:2012:TPE:2335484.2335509} where quotes are made publicly available to a large group of anonymous subscribers,
there exist certain classes of applications where publishers would like to share data, yet only with a closed group of subscribers.

In the aforementioned example of the freight forwarder and its truck fleet, the former intends to publish its geo-location internally and to its clients.
Conversely, a client should only receive location updates of trucks delivering its freight rather than being able to observe the whole fleet movement.
Therefore, we propose an extension to the original publish/subscribe design of content-based routing in the form of \emph{subscription policies}, which allows publishers to exclude certain groups of participants from the reception of certain messages, while still preserving the remaining properties of publish/subscribe systems.

Since publish/subscribe systems have to process large volumes of data, a scalable infrastructure is needed in order to accommodate a large number of publishers and subscribers.
Although the cloud computing paradigm provides an attractive solution, cloud environments must be considered as untrusted environments where the cloud provider cannot be trusted.
To harness the power of scalable infrastructures such as provided by cloud offerings, we propose the use of Intel software guard extensions (SGX)~\cite{Anati2013a,McKeen2013}, which enables the processing of sensitive data in untrusted environments.
Note as well that major Infrastructure-as-a-Service providers (e.g., Microsoft Azure~\cite{azure-confidential} and IBM~\cite{ibm-cloud-data-guard}) are introducing SGX-enabled machines in their commercial offering, suggesting the relevance of our technological choice.

In this paper, we present \projecttitle, a content-based publish/subscribe system that is tailored to applications scenarios that require sharing of information among closed groups of non-anonymous participants.
In order to ensure privacy and confidentiality when operating in untrusted environments, \projecttitle transparently switches to a shielding mode running within Intel SGX enclaves, i.e., each process can run in a separate trusted execution environment (TEE).

The paper makes the following contributions:

($i$) We present the micro-service based architecture of \projecttitle that ensures high scalability to accommodate arbitrary workloads;

($ii$) we propose an extension to the classical publish/subscribe paradigm to allow the sharing of information among closed groups of non-anonymous participants;

($iii$) we provide just-in-time compilation of Python programs inside of enclaves to speed up the execution;

($iv$) we provide the dynamic loading of libraries inside of enclaves to enable the execution of general Python programs---which are often linked with external native libraries; and

($v$) present a performance evaluation when using the shielding mode that utilizes Intel SGX.

The remainder of the paper is organized as follows:
We first provide in Section~\ref{sec:usecase} a detailed description of the use case.
We then detail the architecture of \projecttitle in Section~\ref{sec:architecture}, while implementation details are provided in Section~\ref{sec:implementation}.
In Section~\ref{sec:eval}, we present the performance of our system comparing native vs. trusted execution using Intel SGX.
Finally, we survey related work in Section~\ref{sec:rw} and conclude in Section~\ref{sec:conclusion}.

\section{The \eupro use case}
\label{sec:usecase}

In this section, we describe the use case that motivated the work for \projecttitle.
This use case is taken from the \eupro project~\cite{selis} whose aim is the creation of a common information space and platform for European companies operating in the logistics domain.

The aim is driven by the observation that logistic companies do mostly operate in isolation at the present time and, hence, share only a little or no information with partners and competitors.
This often leads to situations where trucks and vessels return half empty to their hubs although another partner may have freight in need of being delivered on the same route.
Besides a low average utilization, making those trips is often not profitable but also leads to an increased emission of carbon dioxide.

The lack of information exchange does not only cause empty trips on roads and at sea but also causes delivery delays due to unforeseen circumstances often caused by unpredictable conditions such as weather forecasts.
Consider the example of some freight that is first transported through the Mediterranean sea before being loaded onto a truck for an onward journey to its final destination.
As weather conditions are unpredictable, such a vessel may arrive at the port delayed, forcing the truck scheduled for the freight's onward journey to wait and in turn delay the delivery of other freight as well.
Due to the lack of information exchange, such delays cannot be compensated merely by requesting a partner to take over the onward journey if capacity permits.

To overcome those challenges, the \eupro project targets the creation of a common information space as mentioned previously.
The \eupro platform allows logistic companies to share data that might be of interest for other partners, collaborators and clients, for example the GPS position of trucks of a fleet or vessels along with their transported freight containers.
Besides data sharing, the platform also provides additional information and services such as estimated time of arrival (ETA) predictions and key performance indicators (KPIs) tailored to specific participants of the platform.
Data sharing between the different collaborators as well as the different systems in the platform will be enabled through a \emph{content-based publish/subscribe system} acting as information bus in the \eupro platform.

The choice of using a content-based publish/subscribe system rather than a simple topic-based publish/subscribe system such as Kafka~\cite{kreps2011kafka} or RabbitMQ~\cite{rabbitmq} originates from the nature of data shared within the system.
Hence, the data does not only consist of geo-location information for fleets but also of consignments, border control, customs documents, etc. that have specific structures and must be filtered and forwarded based on their content rather than on a simple topic.
Consider for example the GPS position tracking of a truck: for a port authority, only the arrival, i.e., when the truck enters a specific bounding box, may be of interest while a client may want to track the delivery of its freight all the way from the origin up to its destination.
Hence, the GPS updates must be filtered based on complex rules taking into account only specific attributes in publications.

The publish/subscribe paradigm is characterized by a loose coupling, i.e., publishers do not have explicit knowledge about the existence of subscribers and vice versa~\cite{eugster2003many}.
While this scheme works well for most applications and use-cases, the \eupro platform requires a more restrictive message dissemination scheme for the following reason.
Consider a freight forwarder publishing the GPS positions of its truck fleet along with its container contents.
While this information is highly relevant for customs when crossing borders, the container contents itself should not be disclosed to any customer.
Moreover, customers should only be able to subscribe to GPS updates of trucks transporting their goods and not of updates belonging to other customers.
Hence, in \eupro, the message dissemination for certain types of publications must be restricted to a specific group of subscribers or even to a single one.
To achieve this restriction, we propose \emph{a permission filtering stage} and \emph{subscription policies} that can be sent along with publications providing this additional filtering step.
The details for this mechanism are presented in Section~\ref{sec:implementation}.

\eupro is intended to be an open platform accessible for many actors in the logistics domain.
The platform will be running in cloud environments for scalability and to accommodate arbitrary workloads.
However, running applications in cloud environments requires additional measures to ensure privacy and confidentiality.
Although several mechanisms exist to achieve privacy and confidentiality in untrusted environments, such as~\cite{Barazzutti:2012:TPE:2335484.2335509}, we propose the use of Intel SGX as it simplifies the implementation by using plain text matching while providing similar guarantees.

\begin{figure}[t!]
  \centering  
  \includegraphics[scale=0.65]{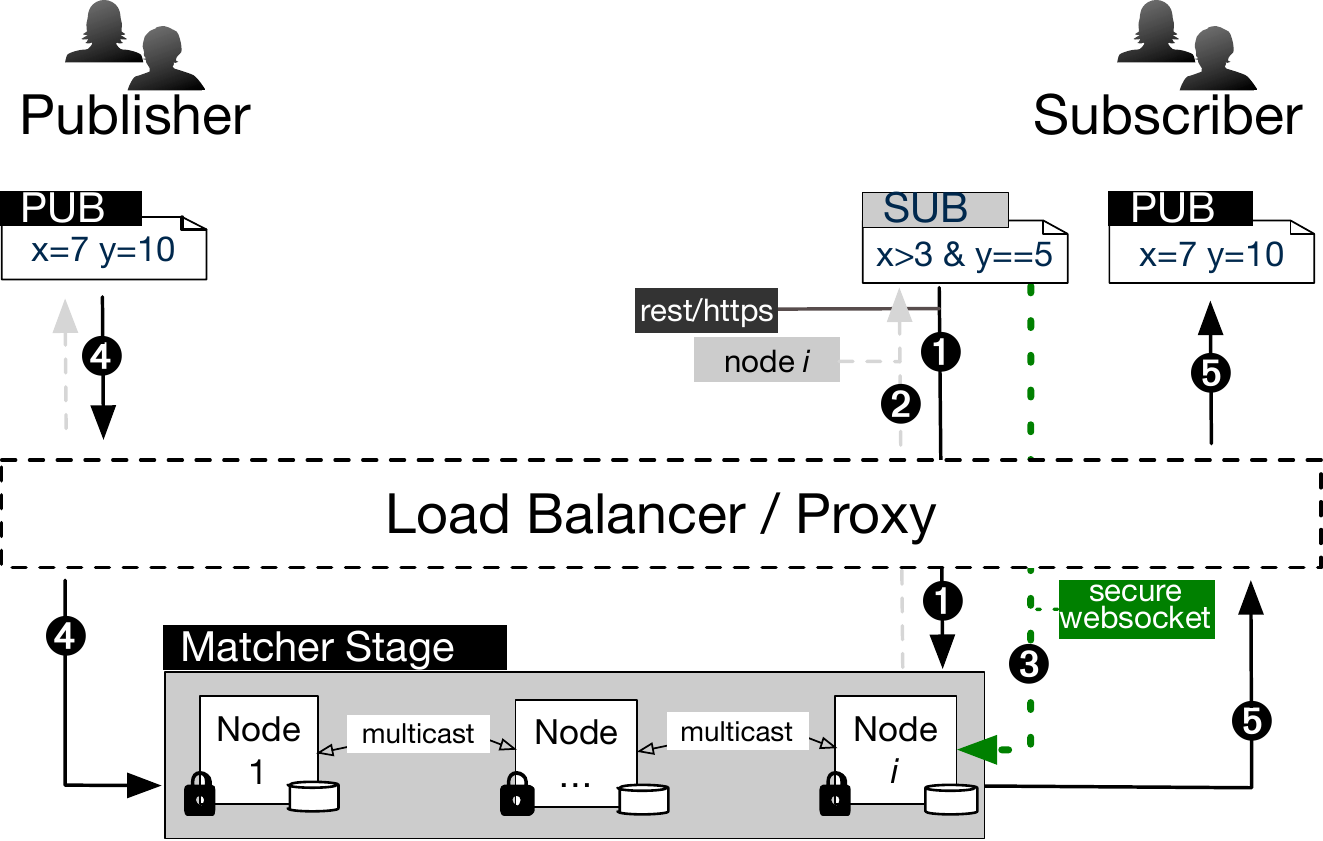}
  \caption{\projecttitle Architecture}
  \label{fig:architecture}
\end{figure}

\section{Architecture}
\label{sec:architecture}

In this section, we present the architecture and design of \projecttitle.
It consists mainly of two components:
a load balancer stage that acts as a proxy and performs a broadcast for publications as well as subscriptions,
and a worker stage with several instances of broker nodes that carry out the actual filtering as depicted in Figure~\ref{fig:architecture}.
Messages transported by the \projecttitle use the JavaScript object notation (JSON).

The stage-based design is inspired by the work presented in~\cite{Barazzutti2013} as it ensures scalability, i.e, the system can accommodate arbitrary workloads.
To achieve scalability, the load balancer stage distributes the load evenly across the matcher instances of the second stage.
We use subscription partitioning, i.e., each matcher instance stores only a subset of the subscriptions.

In the following, we will describe the general message flow for subscriptions and publications using the presented design.
\emph{Subscribers} contact the load balancer stage via a TLS-secured REST endpoint for a secure message exchange (Figure~\ref{fig:architecture}-\ding{202}).
The TLS endpoint, i.e., the load balancer stage, forwards the subscription to the matcher stage, responsible for handling the matching/filtering.
Note that we use subscription partitioning, i.e., each subscription is hashed modulo the number of matcher instances and then forwarded only to one of those matcher instances.
Once the subscription arrives at the matcher instance, the subscription is stored in memory for later matching of publications.
The subscription message contains a set of predicates and attributes used to filter publications.

In addition to storing the subscription at the appropriate matcher node, the load balancer reports back (using the REST response to the subscriber) the ID of the matcher instance that stored the subscription (Figure~\ref{fig:architecture}-\ding{203}).
This allows the subscriber to open a \emph{secure} websocket connection (Figure~\ref{fig:architecture}-\ding{204}) to the load balancer stage indicating from which instance it shall subsequently receive matched publications in a streaming fashion.
A secure websocket is protected by TLS.  

As with subscriptions, \emph{publishers} connect to the system using the same TLS endpoint as shown in Figure~\ref{fig:architecture}-\ding{205}.
However, publications are multicast to all matcher instances since they may all contain matching subscriptions.
The publication is then forwarded to each subscriber that registered a matching subscription using the previously established secure websocket channel (Figure~\ref{fig:architecture}-\ding{206}).

The matcher stage and load balancer/proxy execute inside of SGX enclaves to protect the confidentiality and integrity of the subscriptions and publications.
We attest the matchers as well as the proxy via TLS certificates: the platform ensures that these components only get access to the private keys of their certificates if they are executing inside of enclaves. 

\subsection{Content Matching Stage}

The basic format of the publications in \projecttitle consists of key-value pairs represented in JSON notation.
For example, the following JSON code snippet shows a simple publication of a GPS location update that trucks periodically publish to the system:

\begin{lstlisting}[language=json]
{ "id": "truck-abc",
  "lat": 51.0504,
  "lon": 13.7373}
\end{lstlisting}

Similarly to publications, subscriptions also use a list of key-value pairs.
Subscriptions are additionally equipped with a constraint key-value pair to decide if an attribute, such as \textit{lon} or \textit{lat} shown in the previous example, matches the condition.
For example, the subscription shown below will only match GPS updates whose latitude values fall within the range of 50-60 degrees:

\begin{lstlisting}[language=json]
[{ "key": "lat",
   "val": 50.0,
   "op" : "gt"},
 { "key": "lat",
   "val": 60.0,
   "op" : "lt"}]
\end{lstlisting}

Note that publications only match if all constraints are satisfied, i.e., they use a logical AND.
Hence, in the example given above, two constraints are used to check if both latitude constraints are met.
In case the subscriber opts to express a subscription using a logical OR, it can simply register two individual subscriptions rather than a single one with two constraints.

The application of \projecttitle in the logistic domain showed the necessity for supporting more complex data structures, where the value contains nested objects or list of some values.
To match market needs, \projecttitle allows such publications, however limiting the possibility of defining subscription constraints only to keys with values of primitive types.
The other keys are ignored during the filtering stage.
For the use-cases where the subscription constraints have to be evaluated based on nested values, the publication can be flattened (e.g., using the python \emph{flatten\_json} package).

\subsection{Subscriptions Registration}

Subscriptions are stored on a per user/subscriber basis.
Hence, a subscription is registered using an \emph{authentication hash} derived from a certificate that the subscriber provides when connecting to the system.
The certificate is used to prove the subscriber's identity.
Therefore, only subscribers with a proper certificate can register and remove their own subscriptions.

Internally, subscriptions are stored in a simple hash map, which contains the list of subscriptions for each user.
In order to allow the registration of multiple subscriptions per user, each subscription is also equipped with a unique identifier (\emph{subscription ID}).

After the registration of a subscription, the subscriber must establish a connection to the websocket port of the matcher stage in order to receive publications that match the previously defined subscriptions.
To establish the secure websocket connection, the \emph{authentication hash} and the \emph{subscription ID} are used to identify the subscriber as well as route only messages relevant for this particular subscription on the channel.

\subsection{Permission Filtering Stage}

Although the previously described matching algorithm enables subscribers to filter data only based on publications that are of interest for them, a malicious user could still craft a subscription in such a way that he receives data that a publisher may not necessarily like to share with everyone.

Consider again the example of a truck fleet that regularly publishes GPS locations updates.
While this data may be relevant internally for many systems, only a specific group of users should be able to subscribe and receive GPS locations of a subset of trucks.

Therefore, we introduce a \emph{permission filtering} stage that adds additional constraints in the form of \emph{subscription policies} to narrow down the publication of messages to specific groups of users.
For example, if the GPS locations of a truck identified by its ID \emph{truck-bcd} should be made available only to a specific group \emph{A}, the publisher would install the following constraint at the permission filtering stage:

\begin{lstlisting}[language=json]
[{ "key": "id",
   "val": "truck-bcd",
   "op" : "eq"},
 { "key": "group",
   "val": "A",
   "op" : "eq"}]
\end{lstlisting}

The above filtering rule would apply to any subscriber belonging to group A.
Hence, if a publication passes the regular matching stage, i.e., passing all the content-based constraints that the subscriber expressed, the additional filtering ensures that only GPS location updates of the truck with ID \emph{truck-bcd} will be forwarded to the subscribers of group \emph{A}.
Note that the group field can also be defined in form of a wildcard ``*'', which would apply to any subscriber including anonymous ones.

A common scenario for an application of \emph{subscription policies} is the enforcement of authorization rules in the system, where individual modules, users, companies or departments are allowed to access only limited information due to the privacy constraints.
The permissions of subscribers can be stored in the identity and access management system (IAM), which can be queried later by \projecttitle.
For role-based access control, the configuration file with the permissions can be loaded during startup of \projecttitle. 
 
\section{Implementation Details}\label{sec:implementation}

We designed our system with the goal of building a modular and easy maintainable system that uses as much as possible existing and well maintained technologies.
Our architecture is based on a micro-service approach where the matcher as well as the proxy instances are carried out as REST-based services.
Each of those instances acts as a server and often also as a REST client in order to exchange data between the individual matcher and the proxy nodes.

The micro-service based approach allows us to choose from different technologies as well as to swap out individual components more easily in order to implement/test new features and to investigate performance trade-offs.
In our current implementation, we use GoReplay~\cite{goreplay}, a Go-based proxy/replay server in order to multi-cast publications across a set of matcher instances,
while the matcher instances themselves are implemented in Python using the Flask framework offering a convenient abstraction for REST-based server implementations with TLS support out of the box.
We want to point out that  the TLS connections are terminated inside the SGX enclaves.

In order to provide privacy preserving content-based routing in such a system, we require to protect the confidentiality of all data and the integrity of both the code as well as the data. We use SCONE~\cite{scone}, a framework that supports both Go as well as Python - allowing us to run the different technologies inside Intel SGX enclaves. 

We extended SCONE, however, to support PyPy (just in time Python) as well as dynamic loading of shared libraries during runtime. Dynamic loading of shared libraries is required since many Python modules call functions implemented in shared libraries not loaded at startup.

\subsection{Loading of Libraries at Runtime}

SCONE supports, out of the box, the loading of {\em dynamic libraries} at startup. We have added the support to load dynamic libraries after a program has already been started.

When a program starts, SCONE loads the binary inside of a newly created enclave. Actually, SCONE provides a modified link loader such that SCONE can actually load unmodified binaries inside of enclaves. When a binary is dynamically linked, it also loads all libraries into the enclave the program was linked with. SCONE makes sure that all pages containing the library code are executable.

An enclave has an unique hash that is computed by the CPU when the enclave is set up.  This hash is called {\em MrEnclave}. Every page that is added to an enclave and its permissions, i.e., if the page is readable, writable, and executable, influence {\em MrEnclave}. SCONE also measures, with the help of the CPU, the content of all pages added to an enclave. {\em MrEnclave} therefore reflects both the content of all pages as well as their meta-data (as permissions).

Dynamic libraries loaded during startup are measured, i.e., {\em MrEnclave} will be modified when a library would be modified by an attacker or a failure or any other way. In other words, we can protect the integrity of the code by ensuring that {\em MrEnclave} has a value that the developer expected. SCONE verifies {\em MrEnclave} during startup via remote attestation and ensures that only if {\em MrEnclave} has the expected value, the program can get access to its secrets.

The Python interpreter is typically dynamically linked and hence, shared libraries are loaded during startup into the enclave. Python often loads additional shared libraries when a Python program imports a module. Many Python modules contain also some C implementation to either speed up the execution or since the Python code is only an API to an existing code module. For example, \projecttitle imports the Python {\em threading} library which in turn imports a {\em collection} library which in turns loads a dynamic library {\em \_collections}. Hence, we need to support the loading of libraries after a startup.

To add such a support, we needed to address two problems: 1) inside of enclaves, we cannot change the page permissions, i.e., we cannot make pages into which we load a dynamic library executable, 2) the integrity of these libraries are not protected by {\em MrEnclave} since {\em MrEnclave} is not modified after the initial startup.

We address the first problem as follows. During startup, we make sure that a part of the heap is executable. We can load dynamic libraries into this part and actually be able to execute the functions defined in the library. However, an executable heap introduces a potential security issue since an attacker could inject code there. We can, however, ask the operating system during startup to make pages of the heap non-executable. The operating system can enforce this with the help of the page table. When we load a new shared library, we make this page executable. To exploit executable heap pages, an attacker would need to change the page protection with the help of the operating system first. Note that the next version of SGX will be able to change the permissions of pages after startup. Hence, this problem will soon be addressed by new hardware.  

To address the second problem, we use the file system shield built into SCONE. The file system shield supports the encryption and integrity protection of files. Basically, this shield implements a Merkle tree to verify the integrity of files. We ensure that the integrity of files is checked during the loading into the enclave. Only if dynamic libraries have not been modified, they will be permitted to be loaded into an enclave.

\subsection{Supporting PyPy inside SGX enclaves}

SCONE supports several implementation languages, including Python. 
The standard implementation of SCONE's Python support uses CPython~\cite{cpython}.  
For the development of this work, we added support for PyPy~\cite{PyPy}.
PyPy is a faster alternative to CPython: while CPython simply interprets code, PyPy, which is built over the RPython framework, uses Just-In-Time (JIT) compilation.

As we show in the evaluation section, the support of PyPy resulted in a significant improvement of performance for most Python programs. 
PyPy is broadly applicable: while there are a few micro-benchmarks for which PyPy is slower than CPython, in many longer running benchmarks, PyPy clearly outperforms CPypthon.

To be able to support PyPy, we need to make sure that pages are executable in which PyPy stores the generated code. 
As we explained above, in the current version of SGX, we cannot change the page permissions inside of an enclave. 
We addressed this problem in the same way as above: we support the modification of page permissions by making all heap pages executable and only change the page permissions with the help of the OS.

\subsection{Fault Tolerance and Elasticity}

In the current version of our system, we omitted the implementation of primitives such as checkpointing and
logging~\cite{Elnozahy2002} needed for fault tolerance, as well as state migration techniques to allow the system a seamless scaling 
and adaption to fluctuating workloads.
However, we designed the architecture of our system such that those primitives can be easily added.

First, we use subscription partitioning where the number of partitions can be arbitrarily defined.
Subscription partitioning allows the system to scale with an increasing number of publications and subscriptions, 
as well as providing low overhead fault tolerance as each individual partition can be checkpointed independently.

Second, since partitions are uniquely identifiable using partition IDs and due to the fact that subscriptions and
publications are multi-casted to all instances, we can also provide fault tolerance using active replication~\cite{Martin2011a} by
simply deploying the same partition multiple times onto the system.
Although this replication scheme does not require fault tolerance primitives in place such as checkpointing and logging 
in order to provide high availability, it comes with the disadvantage of delivering matched publications 
multiple times depending on the replication factor chosen for a particular matcher instance.

In order to allow the system to scale elastically during runtime, the previously mentioned active replication scheme can be utilized as in~\cite{Barazzutti2014}.
However, in order to allow the dynamic addition and removal of nodes at runtime, the GoReplay implementation must be slightly
extended to allow the addition and removal of nodes to which the captured traffic will be forwarded to.
Furthermore, the system must be extended with a controller unit that receives performance probes and decides which partition
should be replicated and moved to another spare node if the node the partition is currently running on is close to its saturation point.

Although the previously mentioned primitives allow the system to recover from faults and improve the overall systems availability, 
the system is still prone to message loss and the reception of duplicated messages in the event of a crash of an instance.
This can be circumvented by enforcing message ordering as it ensures that publications and subscriptions are processed in identical
order at every instance.
However, implementing message ordering adds a non-negligible run-time overhead (as authors have shown in~\cite{Martin2011}) which can be omitted if exactly onces processing
semantics are not needed.

\section{Evaluation}\label{sec:eval}

This section reports the results of our evaluation that assess the performance of our system with regards to scalability, as well as overhead when running in shielded mode using Intel SGX hardware.
First, we present our evaluation settings followed by a description of our dataset and data generator used for our benchmarks.

\subsection{Evaluation Settings}

For the experiments, we used a local cluster comprising of 8 nodes, each one equipped with a Intel Xeon E3-1270 v5 CPU with 4 cores running at 3.6 GHz, 8 hyper-threads (2 per core) and 8 MB cache.
Each server has 64 GB of main memory and runs Ubuntu 16.04.3 LTS with Linux kernel version 4.4.
In order to avoid interference of the workload generator and the publish/subscribe instances, we use a separate node equipped with two 14-core Intel Xeon E5-2683 v3 CPUs running at 2 GHz with 112 GB of RAM.
All nodes have a 10 Gb Ethernet NIC connected to a dedicated switch.

\subsection{Workload Description}
In order to mimic the workload used in \eupro, we implemented a workload generator that creates subscriptions based on the subscription profile typically used in \eupro.
In addition to the subscription generator, we also implemented a workload generator for publications as we do not have access to real traces from our project partners.
Publications are generated and submitted through \texttt{wrk2}~\cite{wrk2}, a benchmarking tool for http servers which allows to set predefined data rates to exercise the server.
We leverage \texttt{wrk2}'s \textsc{Lua} scripting interface to extend the generator.
Specifically, we create different publications based on the chosen subscription profile.
Furthermore, we measure the throughput as well as the number of forwarded publications on subscriber side in addition to the publisher interface.
This allows us to verify correctness and measure potential queuing effects.

\subsection{Experiments}
In this section, we will provide an overview of performance measurements of our system.

\subsubsection{Scalability}
In our first experiment, we assess the scalability of the system.
As mentioned previously, we chose Python for the implementation of the matcher instances as it simplifies and reduces the code base as well as takes away the developer's burden of taking care of error prone aspects such as memory management, etc.
Although the use of Python provides many benefits, the global interpreter lock (GIL) limits performance as a python instance is effectively only running in single threaded mode.
As our system is designed to run in a distributed setting, multiple matcher instances can run either on the same node or on different node providing horizontal as well as vertical scale out.

\begin{figure}[t!]
  \centering
        \includegraphics[width=0.85\linewidth]{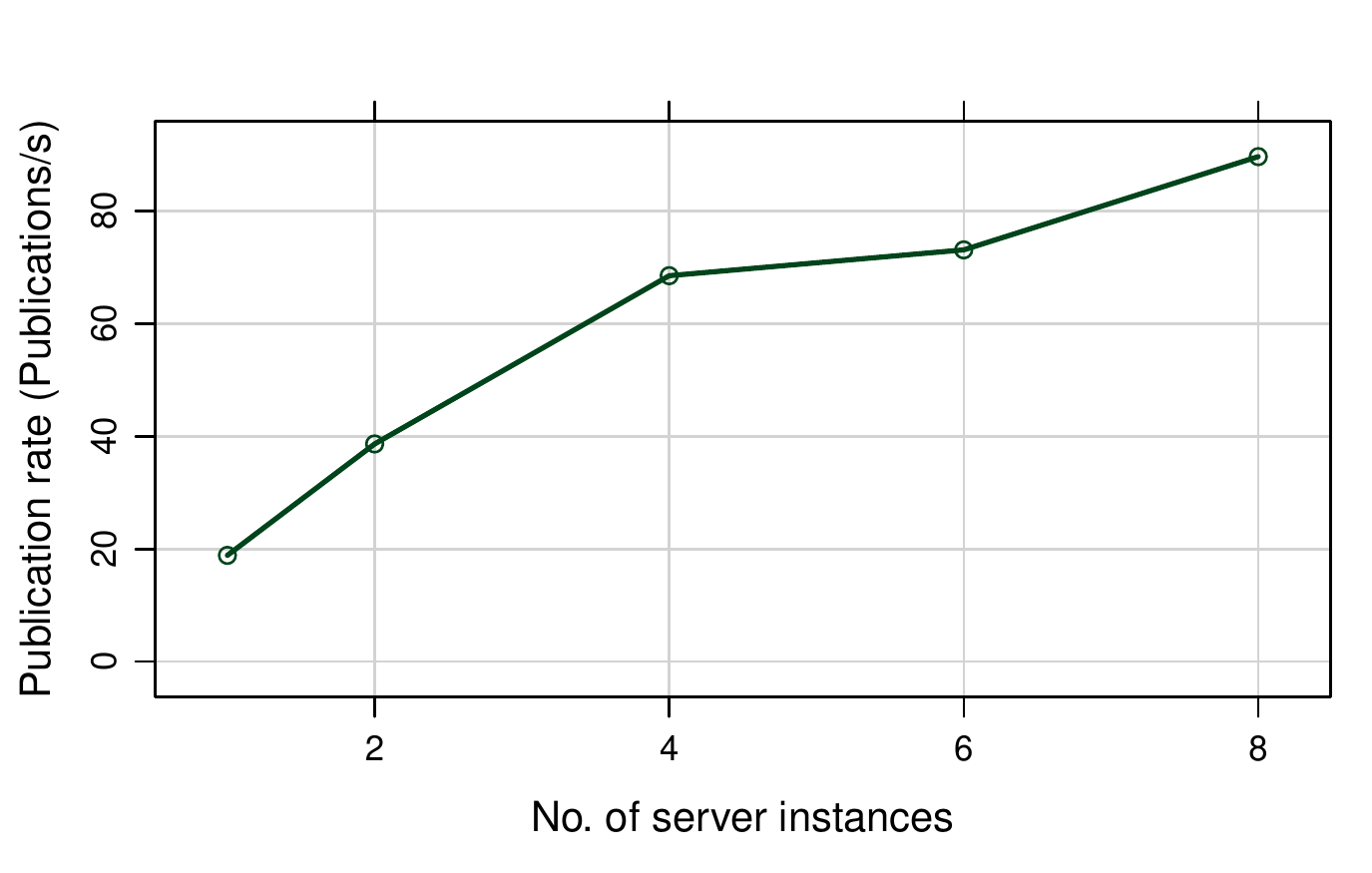}
        \caption{Publication/matching rate with varying number of matcher instances.}
        \label{fig:scaling-instances-tput}
\end{figure}

Figure~\ref{fig:scaling-instances-tput} depicts the scalability of the system. 
For this experiment, we varied the number of matcher instances (from 1-8) but left the number of total subscriptions (256) stored at the system constant.
Note that each subscription comprises 40 attributes that are being matched leading to this relatively low absolute throughput.
Since our system uses subscription partitioning, a higher number of matcher instances leads to less subscription stored in each matcher node, allowing the system to perform the subscription matching in parallel.
As shown in Figure~\ref{fig:scaling-instances-tput}, the initial throughput of 20 publications/s increases as we add more server instances until we reach a peak of approximately 90 publications/s using 8 instances.

\begin{figure}[t!]
\centering
        \includegraphics[width=0.85\linewidth]{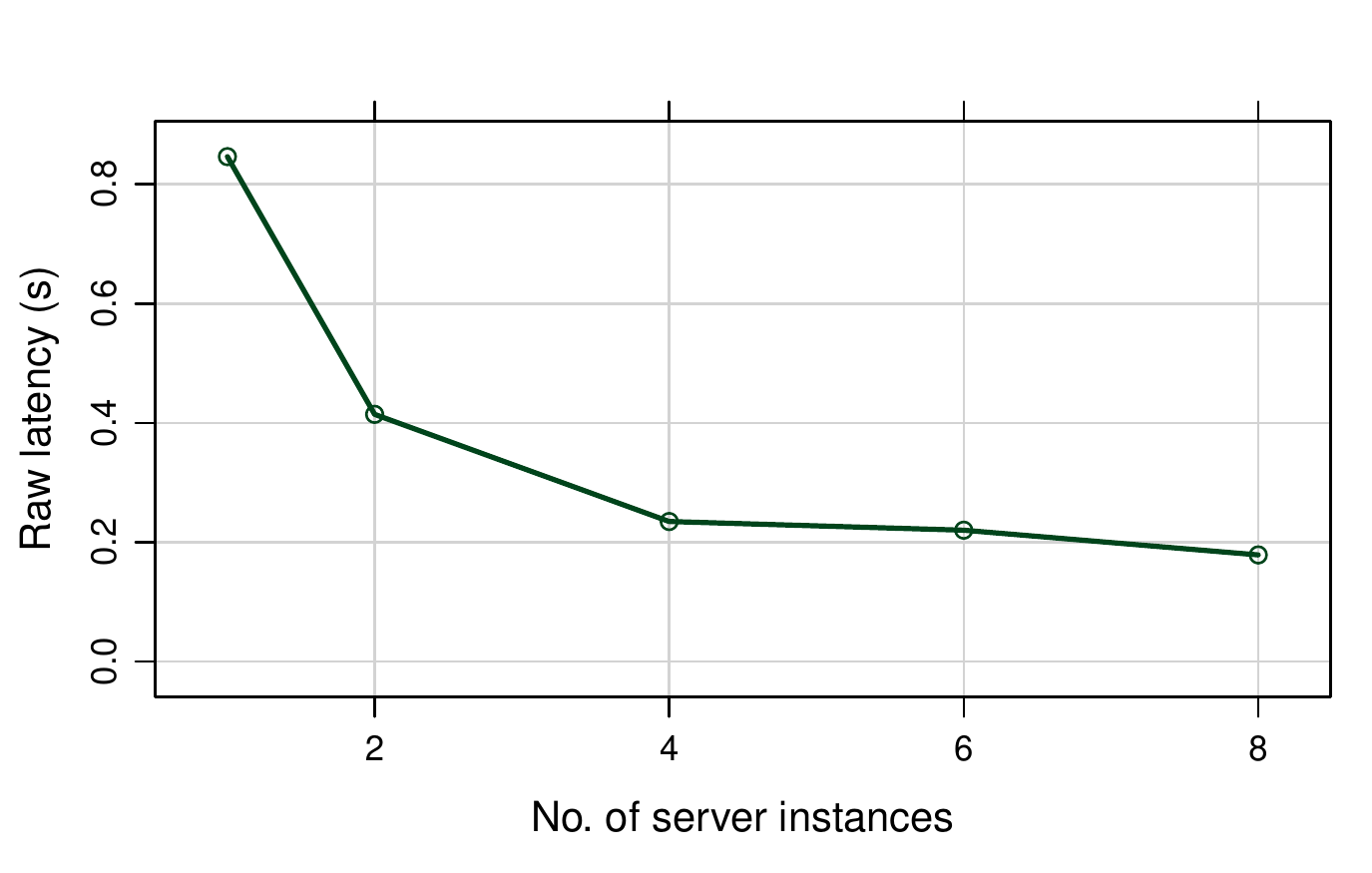}
        \caption{Latency with varying number of matcher instances.}
        \label{fig:scaling-instances-lat}
\end{figure}

The corresponding latency for the same experiment is shown in Figure~\ref{fig:scaling-instances-lat}.
As expected, latency decreases when using more instances as publications are processed in parallel delivering matched publications earlier.

\subsubsection{Overhead Shielding}
In the next experiment, we assess the overhead of using the shielding capabilities to run our system in untrusted environments using Intel SGX.
For this experiment, we used two different versions of Python (CPython and PyPy) and measured throughput and latency of our \projecttitle system when running in native vs. protected/shielded mode.
We furthermore varied the number of subscription stored at the matcher instance.
Note that in our experiment, each subscriber stores only a single subscription, hence, the amount of subscribers is identical to the number of subscriptions stored at the system.

\begin{figure}[t!]
\centering
\includegraphics[width=\linewidth]{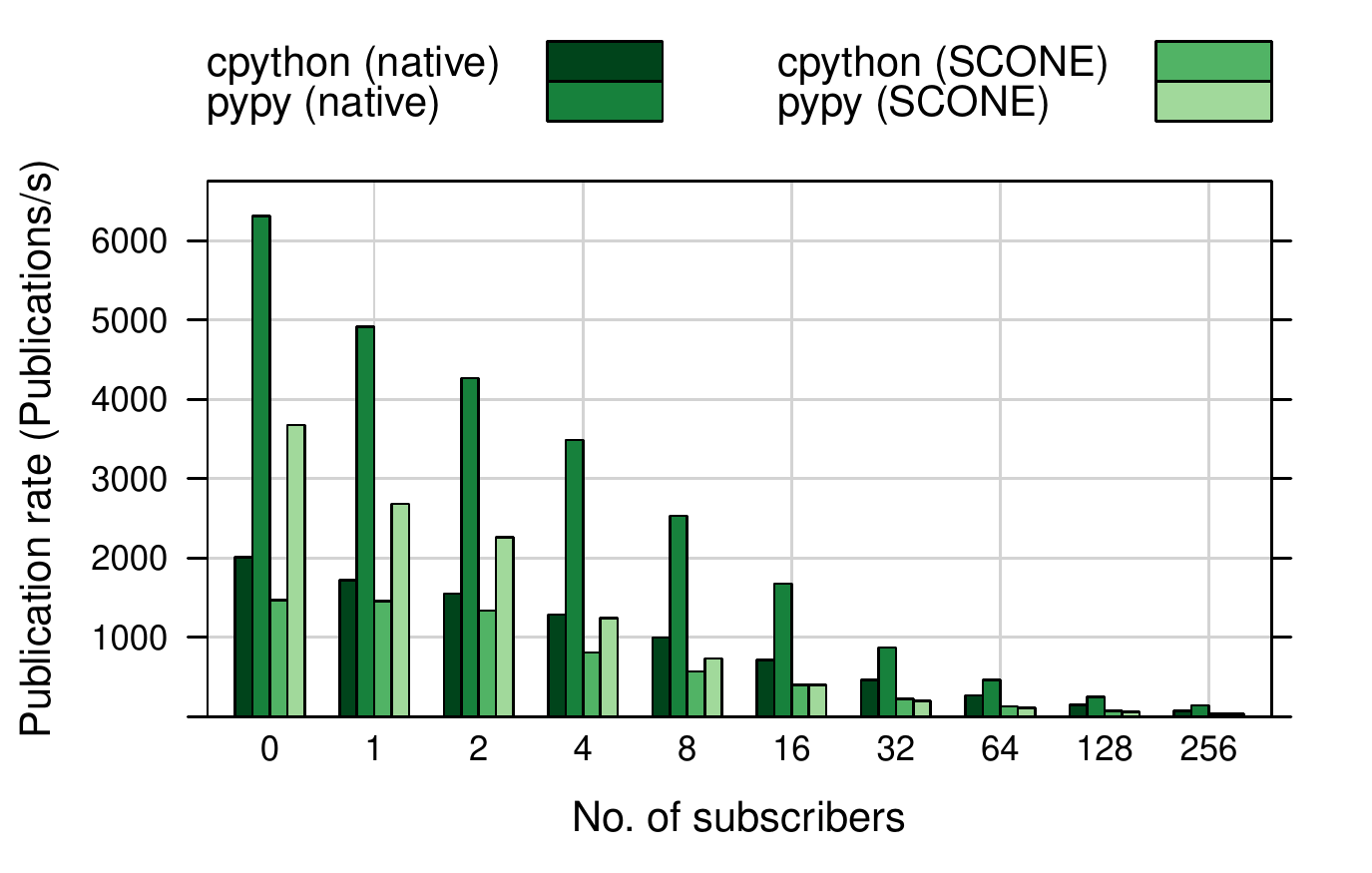}
\caption{Publication/matching rate with varying number of subscribers with and without Intel SGX.}
\label{fig:tput-nat-sgx}
\end{figure}

The results of this benchmark are shown in Figure~\ref{fig:tput-nat-sgx}.
As shown in the graph, throughput decreases with an increasing number of subscriptions stored at the matcher stage.
This is expected as with each incoming publication, more subscriptions must be traversed and checked for a match.
Furthermore, the PyPy implementation outperforms CPython for native as well as shielded execution.
In average, we can observe an overhead for running the system using Intel SGX of approximately $30\%$ in average.
However, when comparing the execution of CPython native versus PyPy using Intel SGX, we can even observe that the shielded version
outperforms the CPython native by about $40\%$ in average.
Interestingly, the performance gain decreases as soon as more subscriptions are stored in the system which recommends the use of more instances.

\begin{figure}[t!]
\centering
        \includegraphics[width=\linewidth]{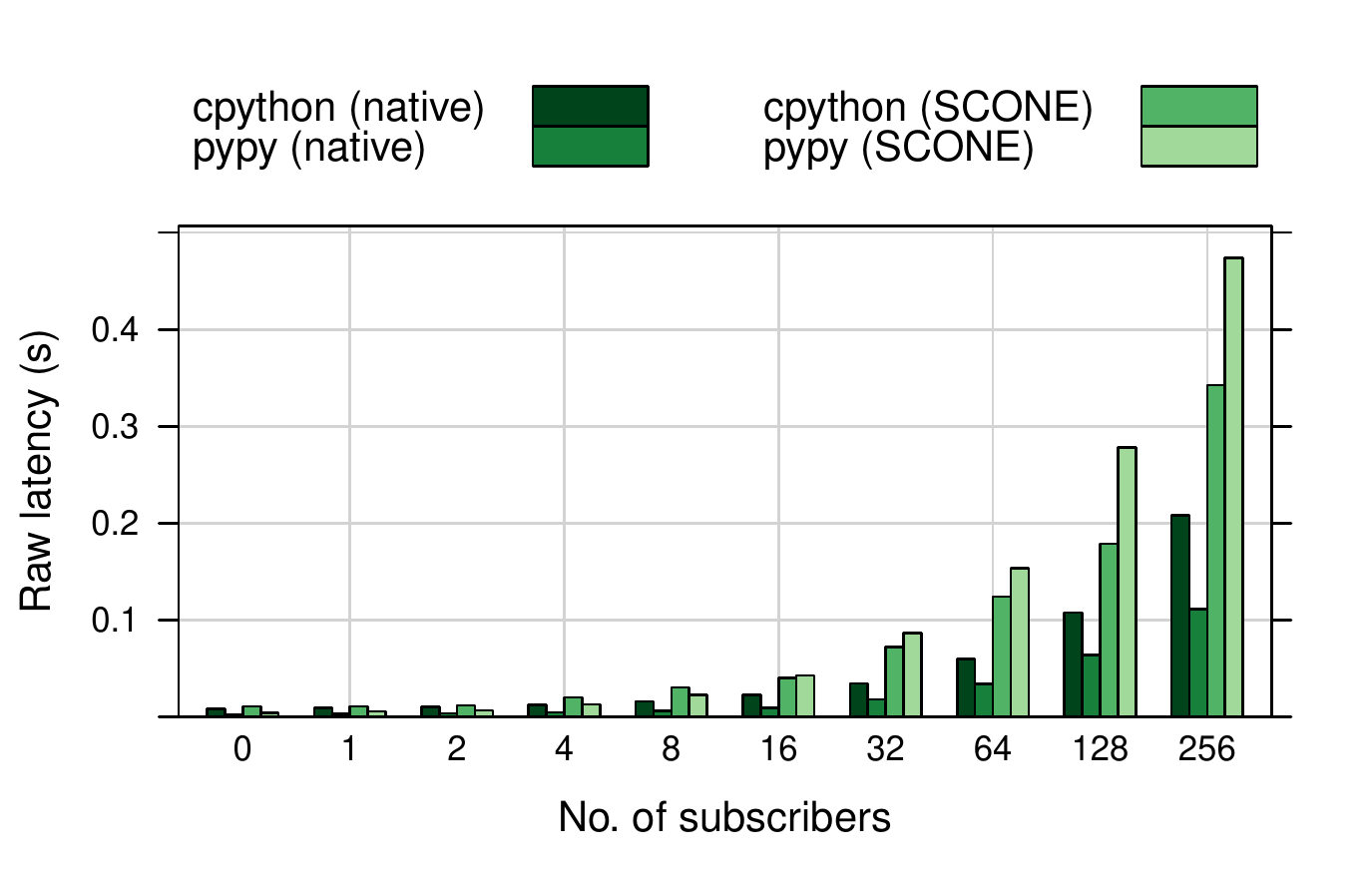}
        \caption{Latency with varying number of subscribers with and without Intel SGX.}
        \label{fig:lat-nat-sgx}
\end{figure}

Again, Figure~\ref{fig:lat-nat-sgx} depicts the evolution of the latency for the same experiment.
As shown in the graph, latency increases with an increasing number of subscriptions as with each incoming publication, 
all subscriptions must be traversed and checked for matches, increasing naturally latency.
Although the SCONE PyPy variant outperforms the CPython native execution, the user pays a price for latency.
In our measurements, we saw a $75\%$ latency increase for CPython (SCONE) in comparison to the native execution,
and 1.2 times the latency when using the PyPy SCONE version.
However, the system still delivers publications in sub-second range, still providing the user with acceptable service.

\begin{figure}[t!]
\centering
        \includegraphics[width=\linewidth]{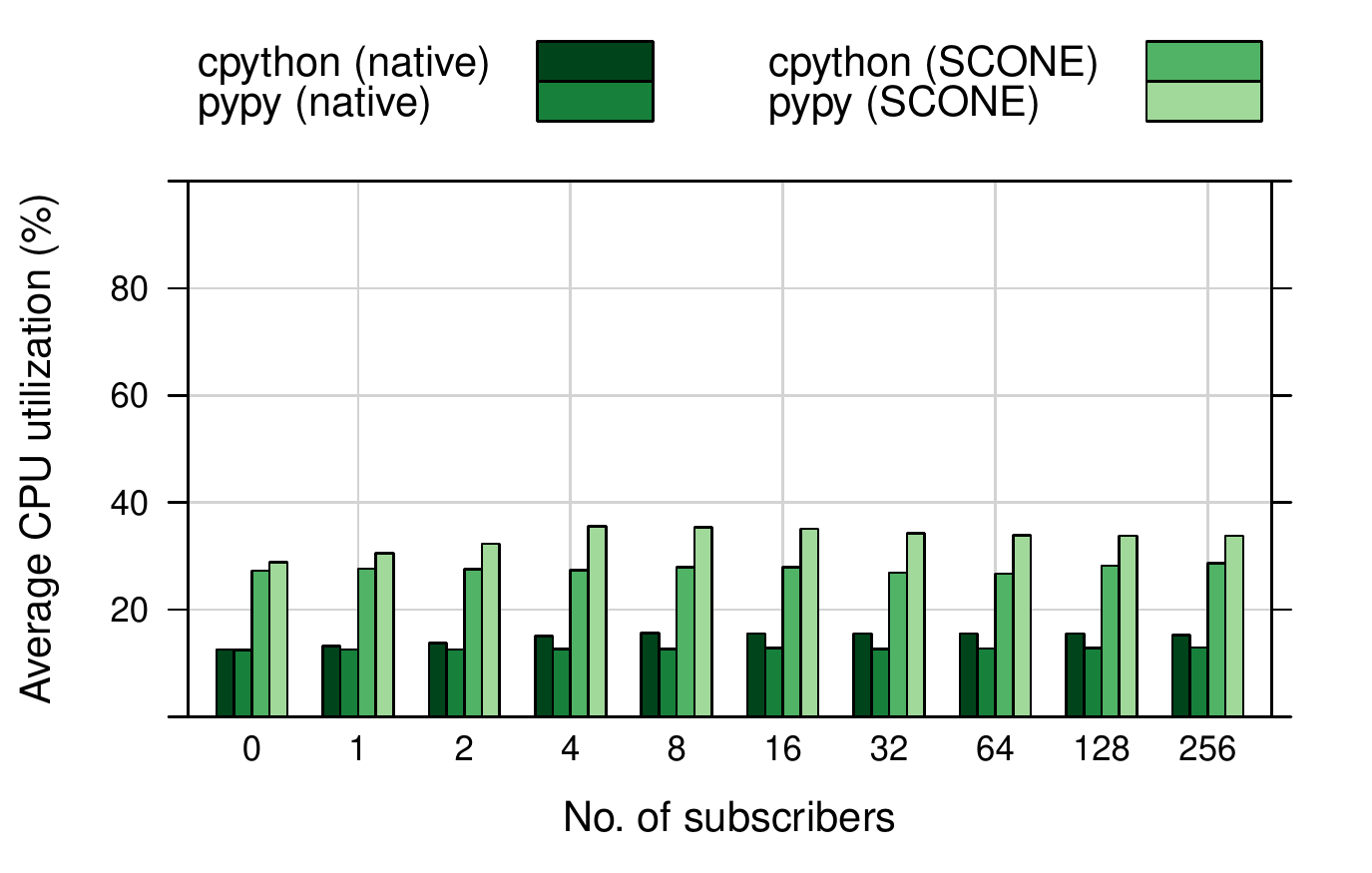}
        \caption{CPU utilization with varying number of subscribers with and without Intel SGX.}
        \label{fig:cpu-nat-sgx}
\end{figure}

In addition to latency and throughput, we also measured the CPU utilization of the system.
The result for this measurement is depicted in Figure~\ref{fig:cpu-nat-sgx}.
As shown in the graph, the CPU utilization stays constant at $12.5\%$ when executing the system in native mode.
This is due to the single threaded execution of our Python-based matcher instance.
Note that the system is equipped with 8 cores, i.e., the full utilization of all course leads to $100\%$.
However, we can see a marginal increase in CPU utilization when more subscriptions are stored and the shielding mode is used.
In general, the shielded version consumes more CPU resources as SCONE uses a worker thread checking for system call returns using a spin lock.
However, as seen in the graph, the CPU utilization does not fully saturate all available CPU resources available on all nodes.
This is due to the Python's global interpreter lock (GIL) leaving the system running in single threaded mode as mentioned previously.

\subsubsection{Vertical Scalability using Workers}

In order to achieve better vertical scalability in addition to the deployment of multiple instances, 
we extended our system such that multiple worker threads perform publication matching in parallel at an instance.
This is achieved by having a worker queue where incoming requests are enqueued and consumed by multiple workers.
Using multiple workers (as processes in Python) allows the matcher instances to make progress despite slow subscribers that consume matched publications
too slowly causing unnecessary back pressure.

\begin{figure}[t!]
\centering
        \includegraphics[width=\linewidth]{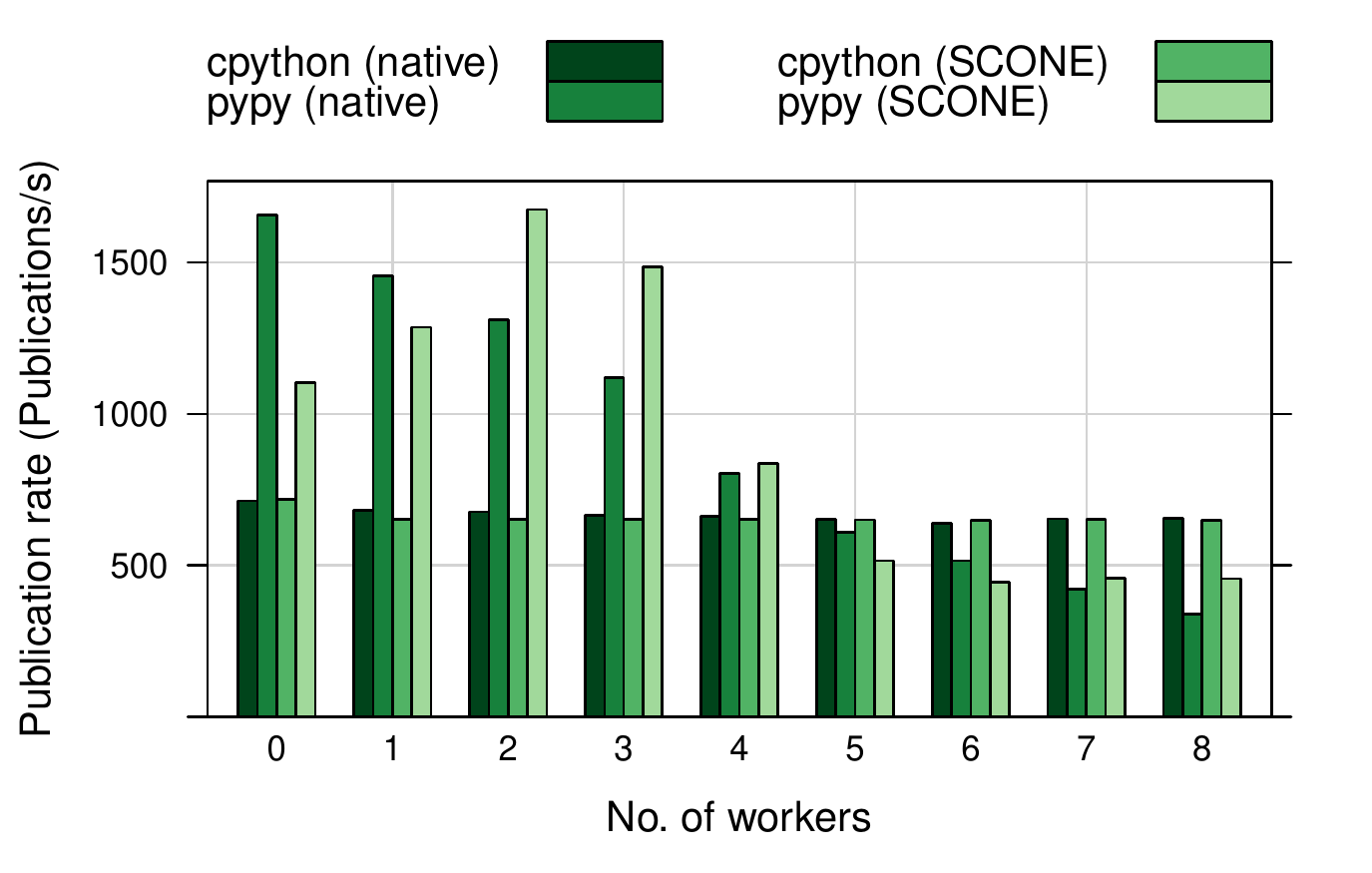}
        \caption{Throughput with varying number of worker threads.}
        \label{fig:tput-workers}
\end{figure}

\begin{figure}[t!]
\centering
        \includegraphics[width=\linewidth]{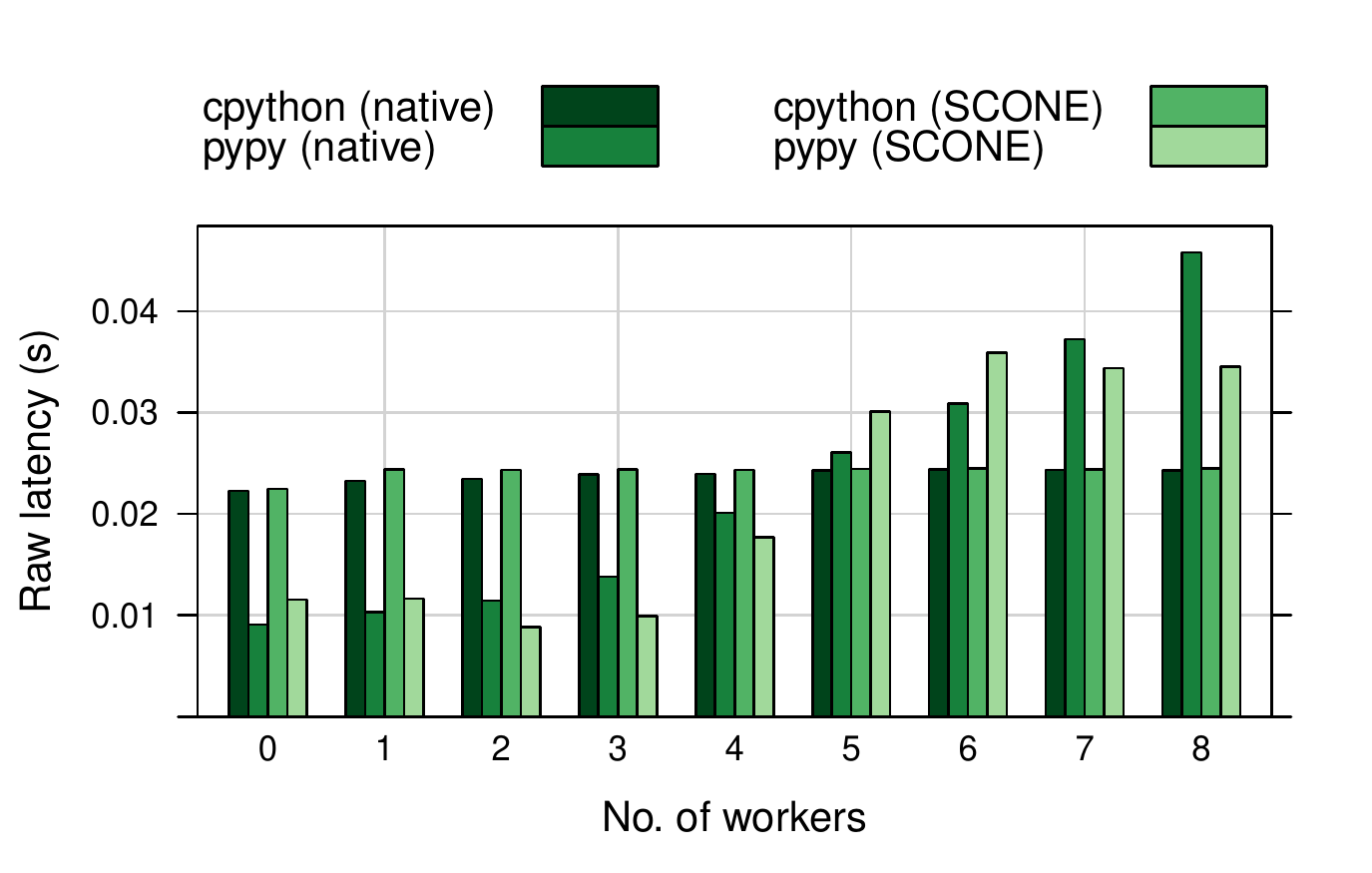}
        \caption{Latency with varying number of worker threads.}
        \label{fig:lat-workers}
\end{figure}

In order to assess the performance for this implementation improvement, we varied the number of worker threads (Python threads) as shown in Figure~\ref{fig:tput-workers}.
As depicted in the graphs, multiple workers lead only to a performance gain in the PyPy shielded variant when using up to 2 workers.
Using more workers, performance decreases as there is too much contention on the single worker queue.
Also latency is reduced as shown in Figure~\ref{fig:lat-workers} as less publishers are blocked due to slow consumers.
Although this approach outperforms the single worker variant (when using PyPy), the implementation lacks of appropriate
back-pressure control causing an increase of enqueued messages if the system is close to its saturation point.

\subsubsection{Permission Filtering}

Contrary to the original design of content-based routing systems, we extended \projecttitle with an additional permission filtering stage.
This stage allows publishers to narrow down the dissemination of certain messages to a closed group of non-anonymous participants desired.

\begin{figure}[t!]
\centering
        \includegraphics[width=\linewidth]{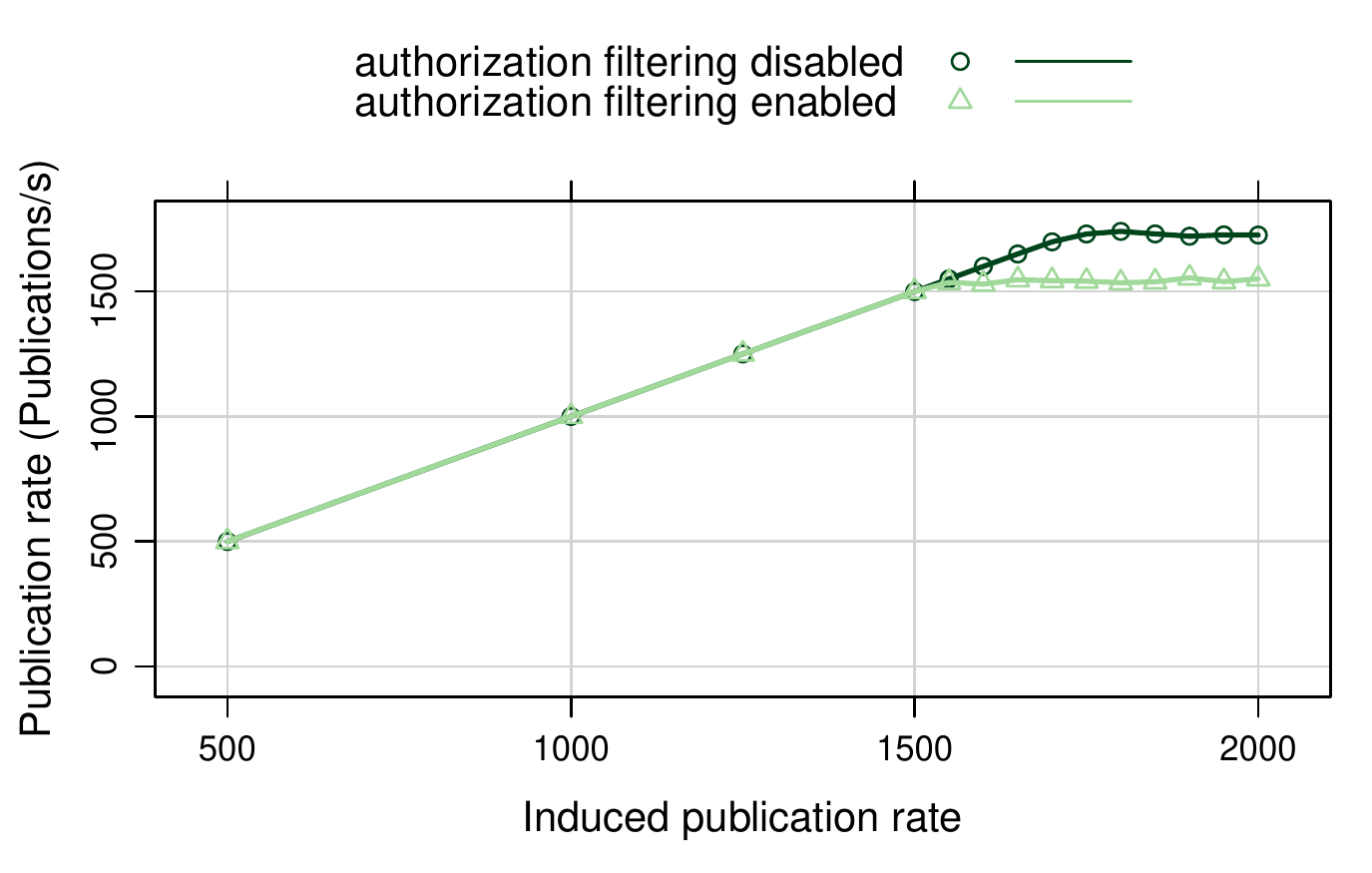}
        \caption{Overhead permission filtering stage in \projecttitle}
        \label{fig:tput-perm}
\end{figure}

Since the permission filtering is applied to every publication that matches content wise, the performance overhead needs to be taken in account as well.
We therefore extended the system such that we can dynamically enable and disable permission filtering.
The results for the overhead introduced through the permission filtering stage is depicted in~\ref{fig:tput-perm}.

In this experiment, we varied the introduced publication rate from 500 publications/s up to the point where the system saturates.
As shown in the figure, the permission filtering adds only as little as $28\%$ overhead when the system reached its saturation point.
If the system is not saturated, there is almost no noticeable overhead when using the permission filtering stage. 
However, in the current setting, we only used a small number of permissions (4).
As more publishers like to restrict message dissemination, we expect higher overheads similar as when increasing the number of subscriptions.

\subsubsection{PyPy \& Shielding Overhead}

In our last experiment, we were interested about the overhead for PyPy and our Intel SGX based shielding for other classes of applications implemented in Python.
The need for the evaluation arises from the \eupro use case where not only information is disseminated through the \projecttitle system
but also KPI and ETA predictions provided in a streaming fashion on top of our system.

\begin{figure*}[htb]
        \includegraphics[width=\linewidth]{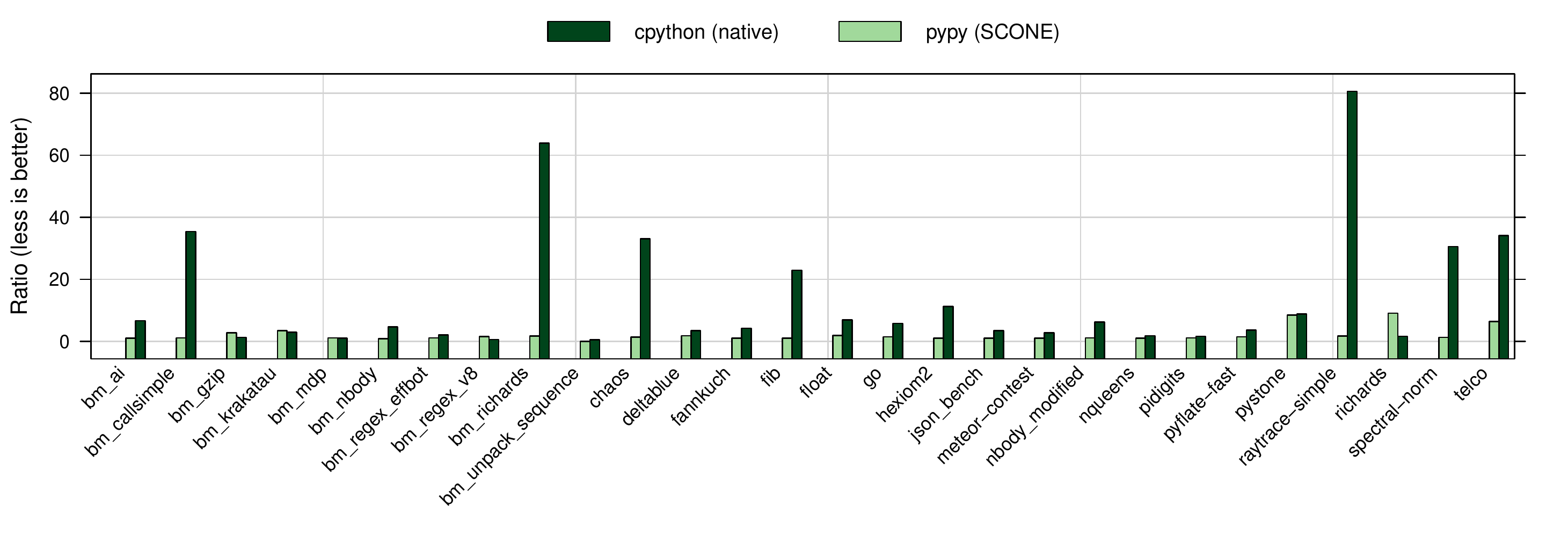}
        \caption{Normalized execution time (against native PyPy) of PyPyScone and native CPython.}
        \label{fig:speedtest}
\end{figure*}

We therefore executed the SpeedTest~\cite{speedpypy} benchmark officially available to assess the performance of PyPy in comparison to CPython as well as its performance when
running using Intel SGX.
The results are depicted in Figure~\ref{fig:speedtest}.

As shown in the graph, the execution time when using PyPy in shielded mode is often on par or even less in comparison with a native execution using CPython.
This clearly advocates for the use of PyPy in conjunction with the shielded execution.
 
\section{Related Work}\label{sec:rw}

There are two main categories of systems that relate with \projecttitle.
In the first category (Section~\ref{ssec:rw:confpubsub}) we report on systems that tackle confidentiality issues in publish/subscribe systems.
In the second one (Section~\ref{ssec:rw:tees}), we briefly overview how TEEs in general (and Intel SGX in particular) are being adopted and used for several different purposes.

To the best of our knowledge, there is only one existing system (SCBR~\cite{pires2016secure}) that efficiently combines the publish/subscribe paradigm with Intel SGX.
SCBR integrates with Intel SGX by running the code of the brokers inside the enclaves, thus preventing an attacker with full control to inspect the messages in transit through the network.
Conversely, \projecttitle prevents unintended messages to being routed to such brokers by means of properly crafted subscription policies, leveraging the mechanisms described earlier.

\subsection{Confidentiality in Publish/Subscribe Systems}\label{ssec:rw:confpubsub}

There is a large body of work on confidentiality issues in publish/subscribe systems~\cite{Onica:2016:CPS:2966278.2940296}.
PP-CBPS is one of the first attempts to provide support for it~\cite{Nabeel:2012:EPP:2295136.2295164}.
It was built on top of the popular SIENA~\cite{carzaniga2001design}.
It relies on heavyweight cryptographic techniques, and in some cases (e.g., equality filtering), it performs the matching within $10\times$ that of the baseline.
As shown in the evaluation, \sys improves greatly on these results by exploiting the hardware support for cryptographic operations offered directly by the TEEs.

PS3~\cite{Pal:2012:PPP:2442626.2442656} exploits Cipher-text Policy Attribute Based Encryption (CP-ABE) and Hidden Vector Encryption (HVE)~\cite{iovino2008hidden} to protect the privacy of subscriber interest and confidentiality of published content.
This system assumes an honest-but-curious threat model as well colluding subscribers.
\sys supports a more powerful threat model, as provided directly by the security guarantees offered by SGX. 

Thrifty~\cite{Barazzutti:2012:TPE:2335484.2335509} uses a pre-filtering stage and containment graphs (inspired by Bloom filters~\cite{bloom1970space}) to implement encrypted matching operators.
This system relies on ASPE~\cite{choi2010privacy}, a cryptographic technique used to implement efficient subscription matching.
\sys does not require the implementation of complex cryptographic primitives, given that confidential data is only managed within the hardware boundaries of the enclaves.

\subsection{Systems Leveraging TEEs}\label{ssec:rw:tees}

Trusted execution environments have gained a lot of traction in the last years.
Several systems and applications have been developed to leverage their additional security guarantees.

TEEs are lately being used to accelerate blockchain-like transaction throughputs~\cite{lind2016teechan}, improving the security of  systems offering anonymity while browsing~\cite{kim2017enhancing}, web searches~\cite{Mokhtar:2017:XRP:3135974.3135987} and more. 

SecureKeeper~\cite{brenner2016securekeeper} implements a privacy-preserving coordination service, preserving the confidentiality of the managed data by exploiting the SGX hardware enclaves.
Similar to \projecttitle, this system does not require the clients to know the keys used by the enclaves to encrypt data toward the data store. 

SecureStream~\cite{Havet:2017:SRM:3093742.3093927} exploits Intel SGX to deploy and run data processing pipelines over untrusted clouds. 
It uses Lua VM~\cite{lua} and ZeroMQ~\cite{zmq} channels to establish secure communications between the stages of the pipeline.
\projecttitle uses Python and secure web-sockets on top of TLS between the matchers, the subscribers and publishers.

Finally, Sampaio et al. built on top of SCBR and consider that publications maybe be restricted to a set of subscribers~\cite{sampaio2017secure}. 
Nevertheless, the approach involves a third-party entity that filters individual subscriptions according to limitations from the publishers. 
Filtering the subscriptions on the publisher side may not be applicable to all cases and may limit scalability.
 
\section{Conclusion}
\label{sec:conclusion}
The publish/subscribe model presents a compelling paradigm for highly scalable communication systems.
This paper introduces the design and evaluation of \sys, a privacy-preserving content-based publish/subscribe system that exploits Intel SGX enclaves.
We implemented a full prototype in Go and Python, also by extending the SCONE container runtime with full support to PyPy.
The experimental evaluation uses a trace generator inspired by a real-world use-case in the domain of logistics. 
The results shows that our system scales well with the number of subscribers and publishers.
Furthermore, our system outperforms CPython's native execution when using shielding in conjunction with PyPy.
Similarly, its CPU overhead is modest if compared against an insecure version.
In future, we plan to add support for fault tolerance and elasticity such that the system can dynamically adapt to fluctuating workloads.

\balance 
\section*{Acknowledgement}
The research leading to these results has received funding from the European Community's Framework Program Horizon 2020 
under grant agreement number 690588 (SELIS), 692178 (EBSIS), 690111 (SecureCloud).

{
\bibliographystyle{IEEEtran}
\bibliography{biblio}
}

\end{document}